\documentclass{kluwer}
\usepackage{graphicx}
\begin{document}
\begin{article}
\begin{opening}
\title{Intermittency of Magnetohydrodynamic Turbulence: Astrophysical Perspective}           
\author{A. Lazarian}
\institute{University of Wisconsin-Madison, Dept. of Astronomy, 475 Charter St., Madison,
WI 53706}


\runningtitle{Intermittency of MHD Turbulence}
\runningauthor{Lazarian}

\begin{abstract} 
 Intermittency 
is an essential property of astrophysical fluids, which demonstrate
an extended inertial range. As intermittency violates self-similarity
of motions, it gets impossible to naively extrapolate the properties
of fluid obtained computationally with relatively low resolution
to the actual astrophysical
situations. In terms of Astrophysics, 
intermittency affects  turbulent heating, momentum transfer, interaction with
cosmic rays, radio waves and many more essential processes. 
Because of its significance, studies of
intermittency call for coordinated efforts from both 
theorists and observers. In terms of theoretical understanding
we are still just scratching a surface of a very rich subject. We have some
theoretically well justified models that are poorly supported by experiments,
we also have She-Leveque model,  which could be vulnerable on theoretical 
grounds, but, nevertheless, is well supported by experimental and 
laboratory data. I briefly discuss a rather
mysterious property of turbulence called ``extended self-similarity''
and the possibilities that it opens for the intermittency research.
Then I analyze simulations of MHD intermittency performed by different
groups and show that their results do not contradict to each other.
Finally, I discuss the intermittency of density,
intermittency of turbulence in the viscosity-dominated regime as well as
the intermittency of polarization of Alfvenic modes. The latter 
provides an attractive solution to account for a slower cascading rate
that is observed in some of the numerical experiments. I conclude by
claiming that a substantial progress in the field may be achieved by
studies of the turbulence intermittency via observations.

\end{abstract}


\end{opening}
\section{What are Intermittency and Extended Self-Similarity?}

Astrophysical fluids are, as a rule, turbulent and magnetized. These 
two properties are closely interrelated. Turbulence
itself can amplify magnetic fields stretching and winding of magnetic
field lines (see Batchelor 1950) and
it is an important ingredient of a more regular
 astrophysical dynamo (see
Mofatt 1978) that is generally believed to be the origin of magnetic
fields on the scale from stars and accretion disks to galaxies (see
a discussion of the problems of the traditional approach in a
review by Vishniac, Lazarian \& Cho 2003 and references therein). 
Magnetic fields, in their turn, constrain motions of ions, which
decreases the diffusivity and increases the Reynolds number $Re$ of the 
flow. The latter is the ratio of the eddy turnover time of a parcel of
gas to the time required for viscous forces to show it appreciably.
Mathematically $Re=LV_L/\nu$, where $L$ and $V_L$ are the scale of
the flow and its velocity, respectively, while $\nu$ is the viscous
diffusivity. Numerically, for most astrophysical astrophysical flows
the Reynolds number is huge, e.g. $Re>10^8$. Its magnetic counterpart
that characterizes to what extend the magnetic field is frozen in the
fluid may be even larger, e.g. $Rm>10^{16}$. Such tremendous $Re$ and 
$Rm$ are not feasible to reproduce in numerical simulations neither now
nor in at any foreseeable future\footnote{Note, that the currently available 3D
simulations for $1024$ cubes have $Re$ and $Rm$ limited by numerical
diffusion of the order of only $10^4$ and those numbers scale linearly
with the cube size.}. Is it possible to understand MHD turbulence
in these circumstances? 

A lot of understanding of the turbulence has been 
achieved by studies of turbulence
self-similarity. This property, which is also called scale-invariance,
implies that
fluid turbulence can be reproduced by the magnification of some part of it.
Take the famous model of incompressible 
Kolmogorov turbulence as an example. In this model
the energy is injected at a large scale $L$ and forms eddies that transfer
the energy to smaller and smaller scales. At the scales where the corresponding
Reynolds number $Re_l\sim lv_l/\nu$ is much larger than unity, the dissipation
over eddy turnover time $t_l\sim l/v_l$ is negligible. As the result, the
energy cascades to smaller and smaller scales without much dissipation, i.e.
$v_l^2/t_l\sim const$, which gives the well-known Kolmogorov power-law
scaling for
the eddies of scale $l$, namely, $v_l\sim l^{1/3}$. The cascade terminates
at the dissipation scale, which provides an {\it inertial range} from
$L$ to $L Re^{-3/4}$, where $Re$ is the Reynolds number corresponding
to the flow at the injection scale $L$.

At the dissipation scales the self-similarity is known to fail with turbulence
forming non-Gaussian dissipation structures as exemplified, e.g. in Biskamp 
(2003). Interestingly enough, present-day research shows that 
self-similarity is not exactly true even along the inertial range. Instead
the fluctuations tend to get increasingly sparse in time and space at 
smaller scales. This property is called {\it intermittency}. Note, that
the power-law scaling does not guarantee the scale-invariance or absence
of intermittency.

From the practical point of view, self-similarity of turbulence simplifies our
description of turbulent phenomena, while intermittency spoils the nice
picture. If intermittency is $Re$ or $Rm$ number dependent phenomenon one
cannot believe to the results of numerical simulations because of the
aforementioned disparity in the numbers. In other words,
 it invalidates naive 
extrapolations of the dynamics of the turbulent fluid
obtained to with computers to real
astrophysical systems.
In terms of astrophysical implications, intermittency
can substantially change heating of interstellar
gas. According to Falgarone et al.  (2005, Falgarone 1999 and references therein) many
chemical endothermal reactions (e.g. formation of CH$^+$) take place
in intensive vortices within interstellar medium. Cho \& Lazarian (2003)
speculated that the mysterious structures in ISM on the AU spatial scale
(see Marscher et al. 1993, Heiles 1997)
can be explained by intermittency of MHD turbulence in the viscosity-damped
regime. Particle diffusion and acceleration may be very different\footnote{Recent changes in understanding of MHD turbulence have induced substantial
shifts in understanding of cosmic ray diffusion and acceleration (see
Yan \& Lazarian (2004), Cho \& Lazarian (2005) and references therein). It 
may happen that intermittency is another vital missing ingredient of the
modern cosmic ray propagation models.} in
intermittent systems (see Honda \& Honda 2005). 
These and many other issues can be settled when we know more
about intermittency of MHD turbulence.

One way to do such studies is to investigate the scaling powers of longitudinal
velocity fluctuations, i.e. $(\delta V)^p$, where 
$\delta V\equiv ({\bf V}({\bf x}+{\bf
r})-{\bf V}({\bf x})) {\bf r}/r$. The infinite set of various powers of
$S^p\equiv \langle (\delta V)^p \rangle$, where $\langle ..\rangle$ denote
ensemble\footnote{In astrophysics spatial or temporal averaging is used.}
 averaging, is equivalent to the p.d.f. of the velocity increments.
For those powers one can write $S^p(r)= a_p r^\xi_p$ to fully characterize
the isotropic turbulent field in the inertial range. While the 
scaling coefficients $a_p$ are given by the values of the function
$S^p$ e.g. at  the injection scale,  the scaling
exponents $\xi_p$ are very non-trivial. It is possible to show that 
for a self-similar flow the scaling exponents are linear function of $n$,
i.e. $\xi_p\sim p$, which for Kolmogorov model $S^1\sim v_l\sim l^{1/3}$ gives
$\xi_p=p/3$. Experimental studies, however, give different results which
shows that the Kolmogorov model is an oversimplified one.

MHD turbulence, unlike hydro turbulence, deals not only with velocity 
fluctuations, but also with the magnetic ones. The intermittencies of the two 
fields can be different. In addition, MHD turbulence is anisotropic as
magnetic field affects motions parallel to the local direction
of ${\bf B}$ very different. This all makes it more challenging to
understand the properties of MHD intermittency 
more interesting.

In spite of the aforementioned limitations of the computational 
approach to studies of
astrophysical turbulence, computers provide the
most cost-effective way of {\it testing} turbulence scalings. The 
problem with both computational and laboratory studies of $S^n$ is
a relatively small inertial range. 

An interesting and yet not understood property of structure functions,
however, helps to extend the range over which $S^p$ can be studied.
Benzi et al. (1993) reported that for hydrodynamic turbulence
the functions $S^p(S^3)$ exhibit much broader power-law range compared
to $S^p(r)$. While for the inertial range a similarity in scaling
of the two functions stem from the Kolmogorov scaling $S^3\sim r$,
the power-law scaling of $S^p(S^3)$ protrudes well beyond the inertial
range into the dissipation range\footnote{In practical terms this means
that instead of obtaining $S^p$ as a function of $r$, one gets $S^p$ as
a function of $S^3$, which is nonlinear in a way to correct for the
distortions of $S^p$.} . This observation shows that the dissipation
``spoils'' different orders of $S$ in the same manner. Therefore there is
no particular need to use the third moment, but one can use any other
moment $S^m\sim r^m$ and obtain a good power law of the function
$S^p\sim (S^m)^{\xi_p/\xi_m}$ (see Biskamp 2003).

Even not understood, the above property of $S^n$ allows to get insight 
into the
scaling of realistic high $Re$ flows with limited $Re$ number simulations
or laboratory experiments. This way the work of Benzi et al. (1993) allowed
much of the progress that we describe below.

\section{What are the measures of intermittency of MHD turbulence?}

Studies of intermittency in incompressible hydro can be traced back to
works of Obukhov (1962) and Kolmogorov (1962). According to the {\it refined
similarity hypothesis} the fluctuations at a scale $l$ can be presented
as $\delta V_l\sim \epsilon_l l^{1/3}$, where $\epsilon_l$ is the 
energy dissipation rate averaged over $l^3$ volume. This model allows for
intermittency and provides $\zeta_p=1/3+\mu_{p/3}$. The corresponding
value of $\mu_p/3$ according to Yaglom (1966) is $\mu_{p/3}=\mu p (p-3)/18$,
where $\mu=2-\xi_6$ and according to numerical simulations in
Vincent \& Meneguzzi (1991) is $\sim 0.2$. With this distribution it is
possible to show that for a typical ISM injection scale of 50pc, 10\%
of the energy of the most energetic events, deposited within a typical
dissipation scale of Alfvenic motions of several hundred of km (the 
Larmor radius of a proton in the cool neutral gas) is localized in just
$10^{-2}$\% of the volume! 

However, a more successful model to reproduce both experimental hydro data
and numerical simulations is She-Leveque (1994) model. According to Dubrulle 
(1994) this model can be derived assuming that the energy from large scale
is being transferred to $f<1$ less intensive eddies and $1-f$ of more intensive
ones.   
The scaling relations suggested by She \& Leveque (1994) 
related $\zeta_p$ to the scaling of the velocity $V_l\sim l^{1/g}$,
the energy cascade rate $t_l^{-1}\sim l^{-x}$, and the co-dimension of the
dissipative structures $C$:
\begin{equation}
\zeta_p={p\over g}(1-x)+C\left(1-(1-x/C)^{p/g}\right).
\label{She-Leveque}
\end{equation}
For incompressible turbulence these parameters are $g=3$, $x=2/3$, 
and $C=2$, implying that
dissipation happens over 1D structures (e.g. vortices). There are theoretical
arguments against the model (see Novikov 1994), but so far the
She-Leveque scaling is the best for reproducing the intermittency of
incompressible hydrodynamic turbulence.

In their pioneering study
Muller \& Biskamp (2000) applied the She-Leveque model to incompressible 
MHD turbulence and 
attracted the attention of the MHD researchers to this tool. They
used Elsasser variables and claimed  that their results are consistent 
with dissipation within 2D structures (e.g. 2D current sheets). The consequent
study by Cho, Lazarian \& Vishniac (2002a) used velocities
instead of Elsasser variables and provided a different
answer, namely, that the dimension of dissipation structures is
the same as in incompressible hydro, i.e.
the dissipation structures are 1D. The difference between the two
results was explained in Cho, Lazarian \& Vishniac (2003, henceforth
CLV03). They noted that, first of all, 
the measurements in Muller \& Biskamp (2000) were done
in the reference frame related to the {\it mean} magnetic field, while
the measurements in Cho, Lazarian \& Vishniac (2002a) were done in
the frame related to the {\it local} magnetic field. We believe that
the latter
 is more physically motivated frame, as it is the local magnetic field
is the field that is felt by the eddies. It is also in this reference frame
that the scale-dependent anisotropy predicted in the Goldreich-Shridhar (1995, henceforth GS95)
model is seen. Computations in CLV03 confirmed that the dissipation structures
that can be identified as velocity
 vortices in the local magnetic field reference frame can also
be identified with
two dimensional sheets in terms of Elsasser variables 
in the mean magnetic field reference frame.
This, first of all, confirms a mental picture where motions perpendicular
to magnetic field lines are similar to hydrodynamic eddies. More importantly,
it sends a warning message about the naive interpretation
of the She-Leveque scalings in the MHD turbulence.  

Intermittency in compressible MHD turbulence was discussed in Boldyrev (2002)
who assumed that the dissipation there happens in shocks and therefore
the dimension of the dissipation structures is 2. The idea of the dominance of
shock dissipation does not agree well with the numerical simulations in
Cho \& Lazarian (2002, 2003, henceforth CL03), where the dominance of the vortical motions
in {\it subAlfvenic} turbulence (i.e. magnetic pressure is larger than the
gaseous one)  was reported. 
Nevertheless, numerical simulations
in Padoan et al (2003) showed that for {\it superAlfvenic} turbulence 
(i.e. magnetic
pressure is less than the gaseous one) the dimension of the
dissipation structures was gradually changing from one to somewhat
higher than two as the
Mach number was increasing from 0.4 to 9.5. The very 
fact that the superAlfvenic
turbulence, which for most of the inertial scale resolvable by simulations
does not have a dynamically important magnetic field is different from
subAlfvenic is not surprising. The difference between the results
in Padoan et al. (2003) at low Mach number and the incompressible runs in
Muller \& Biskamp (2000) deserves a discussion, however. First of all,
the results in Padoan et al. (2003) are obtained for the velocity, while
the results in Muller \& Biskamp (2000) are obtained for the Elsasser.
CLV03 has shown that the magnetic field and velocity have different
intermittencies. Indeed, it is clear from Fig.~1 that 
$\zeta^{\rm magnetic}<\zeta^{\rm velocity}$  which means that magnetic
field is more intermittent than velocity. An interesting feature of
superAlfvenic simulations in Fig. 1 is that the velocity follows the
She-Leveque (1994) hydro scaling with vortical
dissipation, while magnetic field exhibits a pronounced dissipation
in current sheets. Both features are expected if magnetic field is
not dynamically important and the turbulence stays essentially hydrodynamic.
We also see that the dynamically important magnetic field does changes
the intermittency. The flattening of magnetic field scaling is pronounced
in Fig.~1.

\begin{figure*}
  \includegraphics[width=0.3\textwidth]{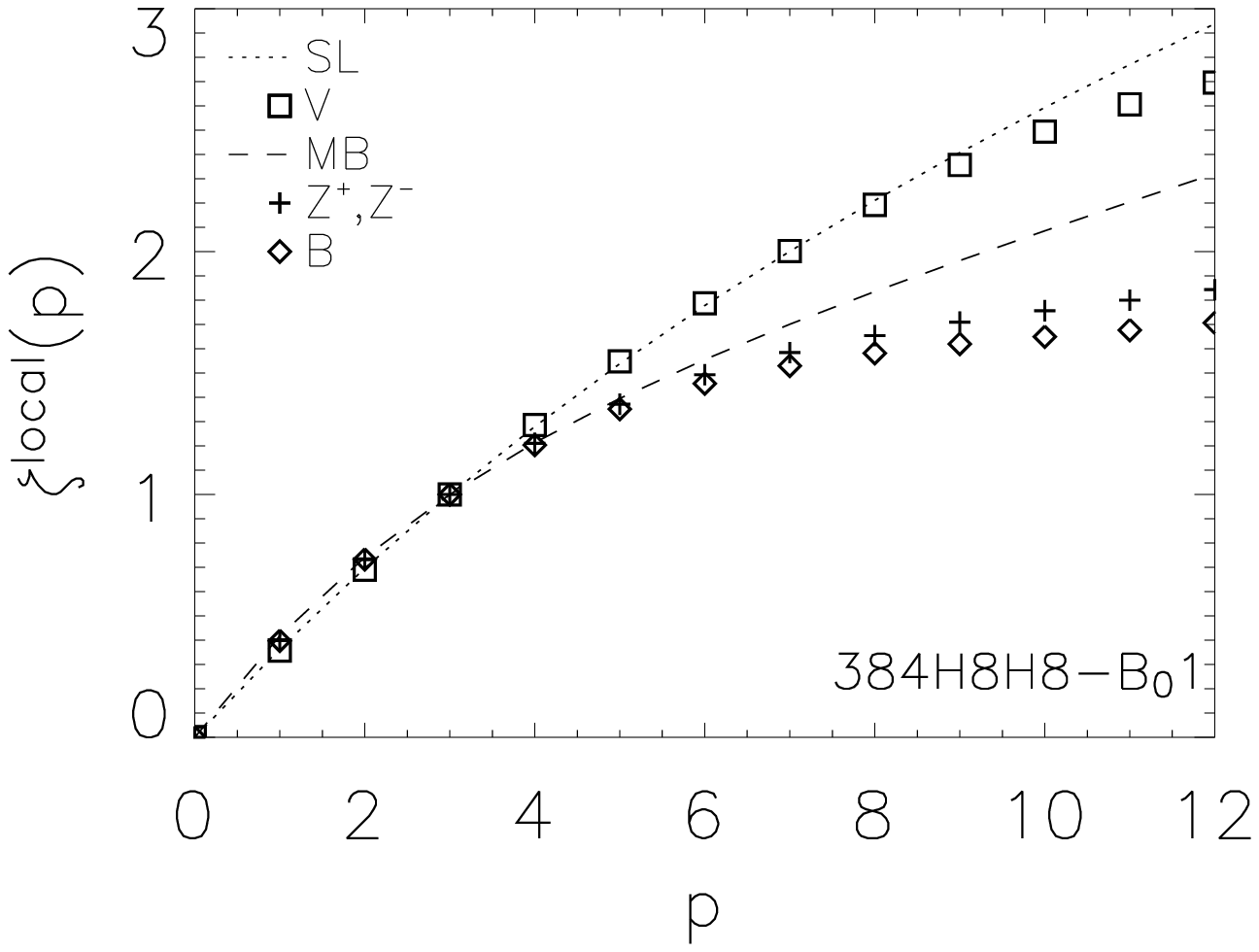}
\hfill
  \includegraphics[width=0.3\textwidth]{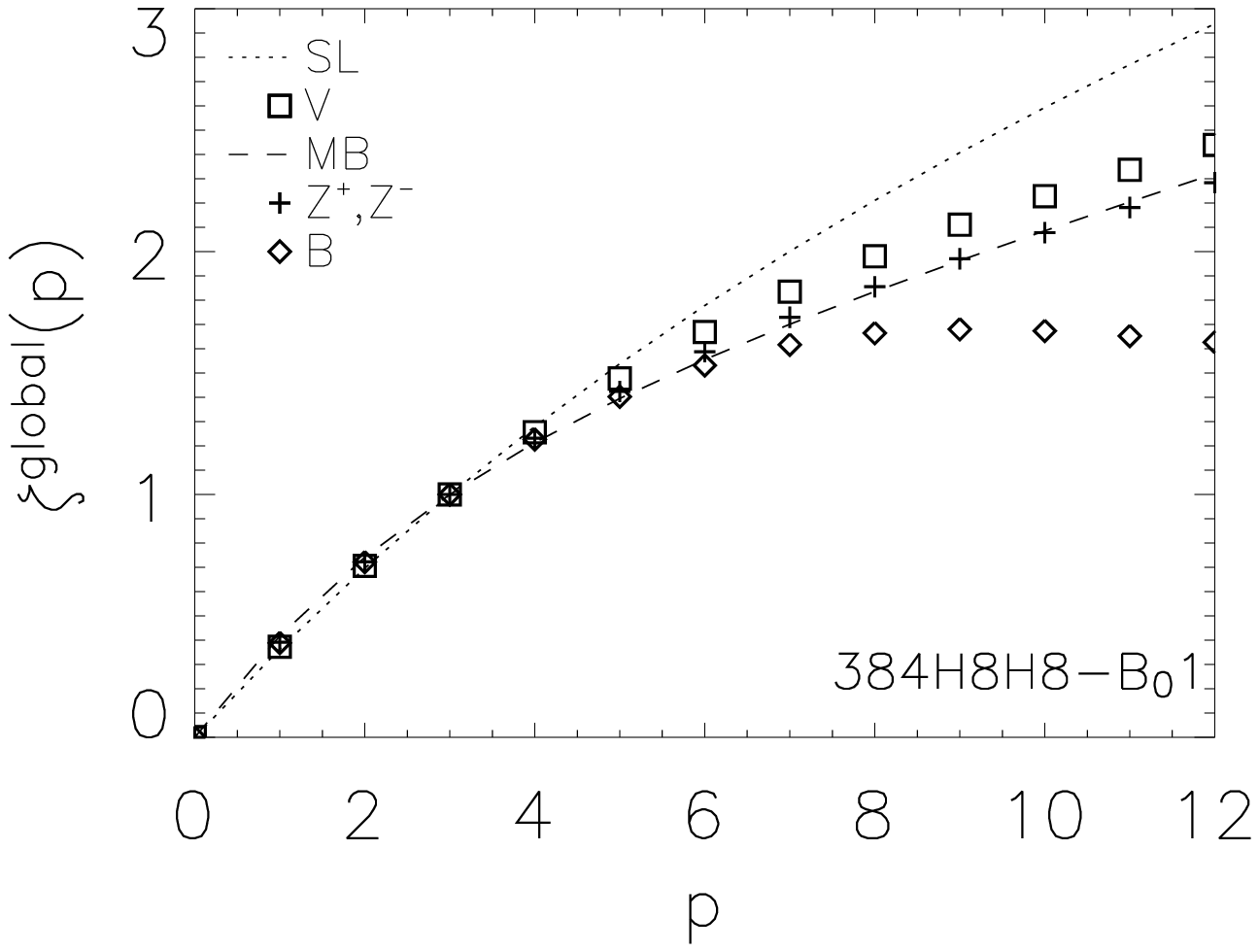}
\hfill
  \includegraphics[width=0.3\textwidth]{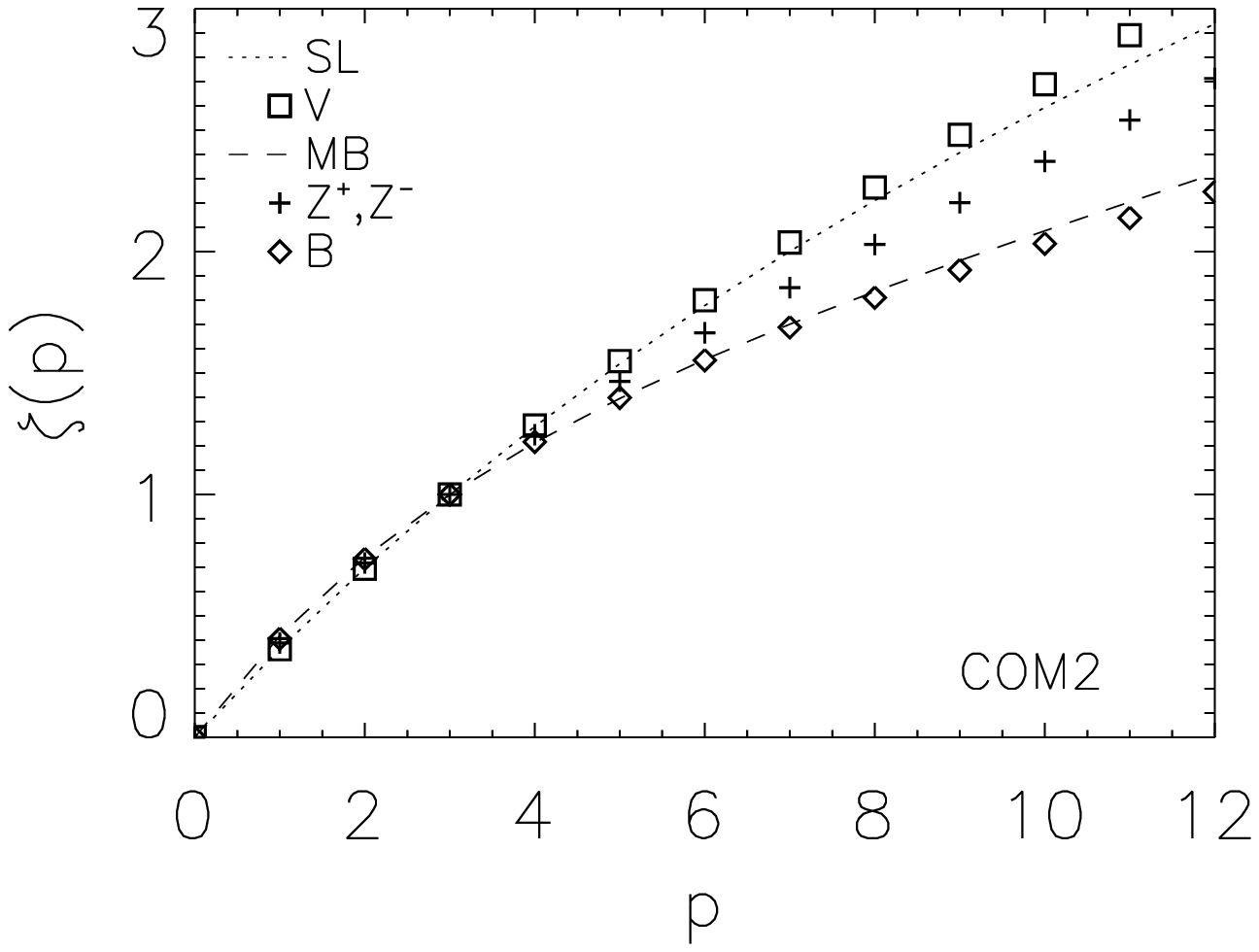}
  \caption{ {\it left panel}: Intermittency  
        exponents for incompressible MHD turbulence
        in perpendicular directions 
        in the local frame.  
        The velocity exponents show a scaling similar to the 
        She-Leveque model. 
        The magnetic field shows a different scaling.
 {\it central panel}: Intermittency 
        exponents for incompressible MHD turbulence
        in the global frame. 
        Note that the result for $z^{\pm}$ is very similar to  
        the M\"{u}ller-Biskamp model.
{\it right panel}: Intermittency exponents for
superAlfvenic compressible 
turbulence in the global frame. From CLV03
}
\label{fig_components}
\end{figure*}

\section{What are the intermittencies of Alfven, slow and fast components
of MHD cascade?}

A further studies of intermittencies was done in Kowal \& Lazarian (2006).
There we used a decomposition of the MHD turbulence into Alfven, slow
and fast modes from Cho \& Lazarian (2002)(see Fig.~2). How physical is 
this decomposition? If the coupling between the modes is strong in MHD turbulence one
cannot talk about three different energy cascades. 
Indeed, the compressible MHD turbulence is a highly non-linear phenomenon
and it has been thought that 
Alfven, slow and fast modes 
are strongly coupled. Nevertheless,
one may question whether this is true.
A remarkable feature of the GS95 model is that
Alfven perturbations cascade to small scales over just one wave
period, while the other non-linear interactions require more time.
Therefore one might expect
that the non-linear interactions with other types of waves
should affect Alfvenic cascade only marginally. 
Moreover, since the Alfven waves are incompressible, the properties
of the corresponding cascade may not depend on the sonic Mach number.

\begin{figure*}
  \includegraphics[width=0.90\textwidth]{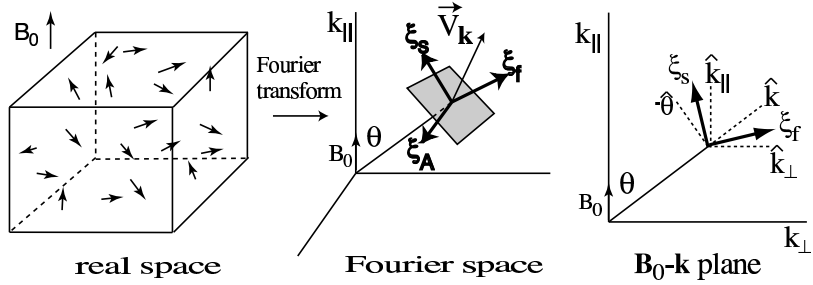}
  \caption{
      Separation method. We separate Alfven, slow, and fast modes in Fourier
      space by projecting the velocity Fourier component ${\bf v_k}$ onto
      bases ${\bf \xi}_A$, ${\bf \xi}_s$, and ${\bf \xi}_f$, respectively.
      Note that ${\bf \xi}_A = -\hat{\bf \varphi}$. 
      Slow basis ${\bf \xi}_s$ and fast basis ${\bf \xi}_f$ lie in the
      plane defined by ${\bf B}_0$ and ${\bf k}$.
      Slow basis ${\bf \xi}_s$ lies between $-\hat{\bf \theta}$ and 
      $\hat{\bf k}_{\|}$.
      Fast basis ${\bf \xi}_f$ lies between $\hat{\bf k}$ and 
      $\hat{\bf k}_{\perp}$.  From CL03
}
\label{fig_separation}
\end{figure*}

The generation of compressible motions 
(i.e. {\it radial} components in Fourier space) 
{}from Alfvenic turbulence
is a measure of mode coupling.
How much energy in compressible motions is drained from Alfvenic cascade?
According to closure calculations (Bertoglio, 
Bataille, \& Marion 2001; see also Zank \& Matthaeus 1993),
the energy in compressible modes in {\it hydrodynamic} turbulence scales
as $\sim M_s^2$ if $M_s<1$.
CL03 conjectured that this relation can be extended to MHD turbulence
if, instead of $M_s^2$, we use
$\sim (\delta V)_{A}^2/(a^2+V_A^2)$. 
(Hereinafter, we define $V_A\equiv B_0/\sqrt{4\pi\rho}$, where
$B_0$ is the mean magnetic field strength.) 
However, since the Alfven modes 
are anisotropic, 
this formula may require an additional factor.
The compressible modes are generated inside the so-called
GS95 cone, which takes up $\sim (\delta V)_A/ V_A$ of
the wave vector space. The ratio of compressible to Alfvenic energy 
inside this cone is the ratio given above. 
If the generated fast modes become
isotropic (see below), the diffusion or, ``isotropization'' of the
fast wave energy in the wave vector space increase their energy by
a factor of $\sim V_A/(\delta V)_A$. This  results in
\begin{equation}
  \frac{\delta E_{comp}}{\delta E_{Alf}}\approx \frac{\delta V_A V_A}{V_A^2+c_s^2},
\label{eq_high2}
\end{equation}
where $\delta E_{comp}$ and $\delta E_{Alf}$ are energy
of compressible  and Alfven modes, respectively.
Eq.~(\ref{eq_high2}) suggests that the drain of energy from
Alfvenic cascade is marginal
when the amplitudes of perturbations
are weak, i.e. $(\delta V)_A \ll  V_A$. Results of numerical
calculations
shown in Cho \& Lazarian (2002) support these theoretical considerations.
This justifies\footnote{A claim in the literature is that a 
strong coupling of incompressible and compressible motions is
required to explain simulations that show fast decay of MHD turbulence.
There is not true. The incompressible motions decay themselves 
in just one Alfven crossing time.} our treating modes separately.

\begin{figure*}
  \includegraphics[width=0.3\textwidth]{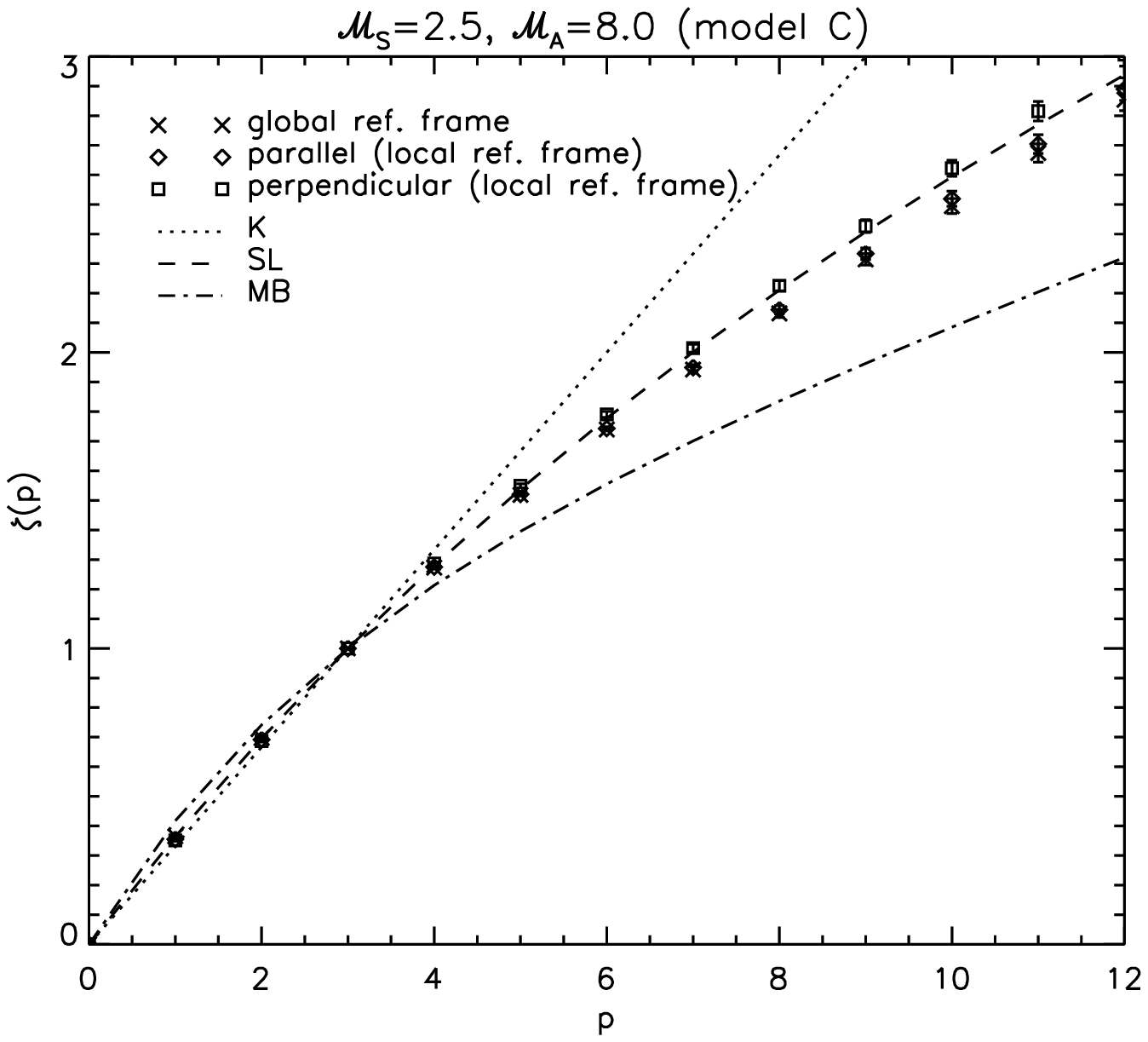}
\hfill
  \includegraphics[width=0.3\textwidth]{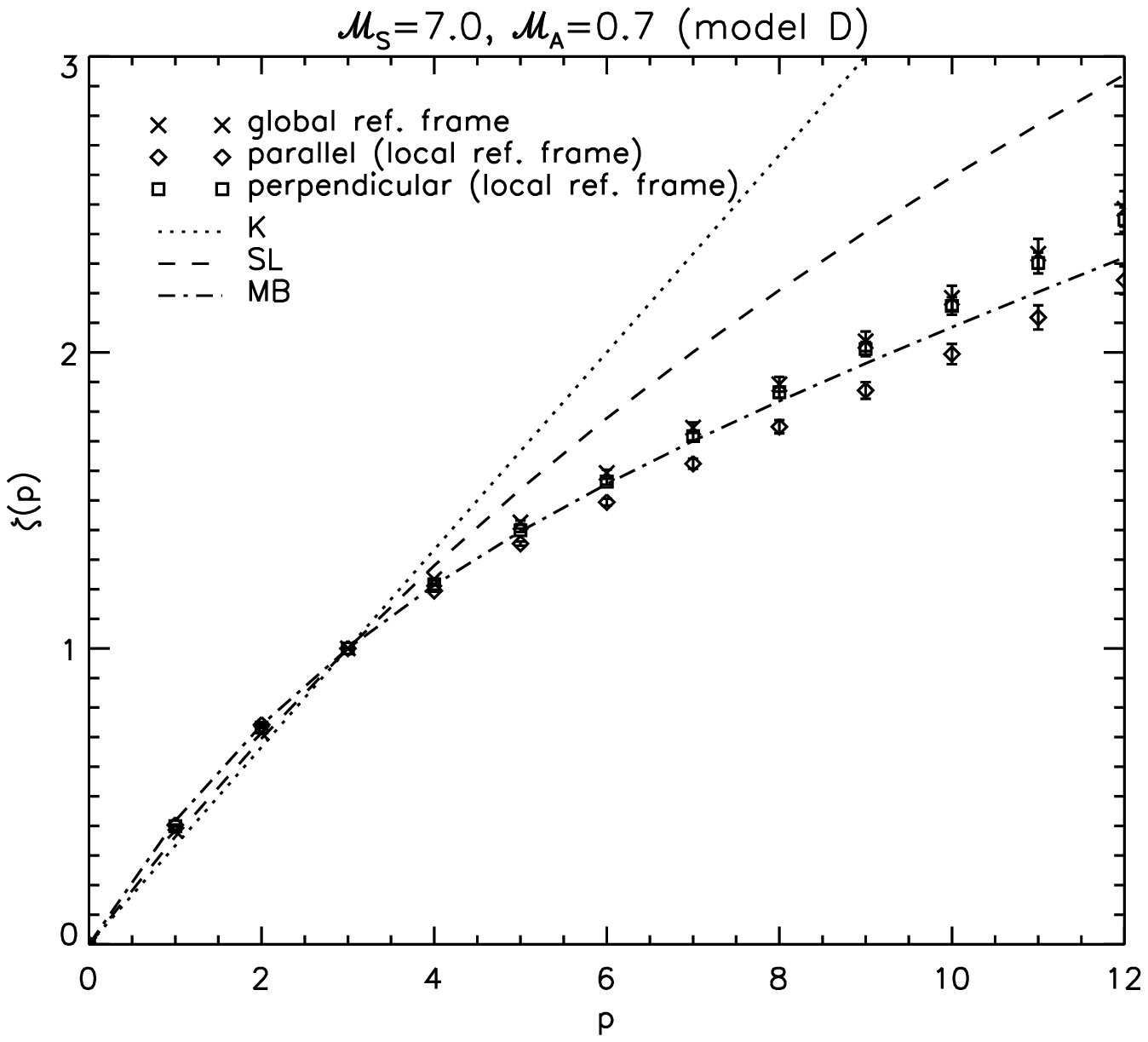}
\hfill
  \includegraphics[width=0.3\textwidth]{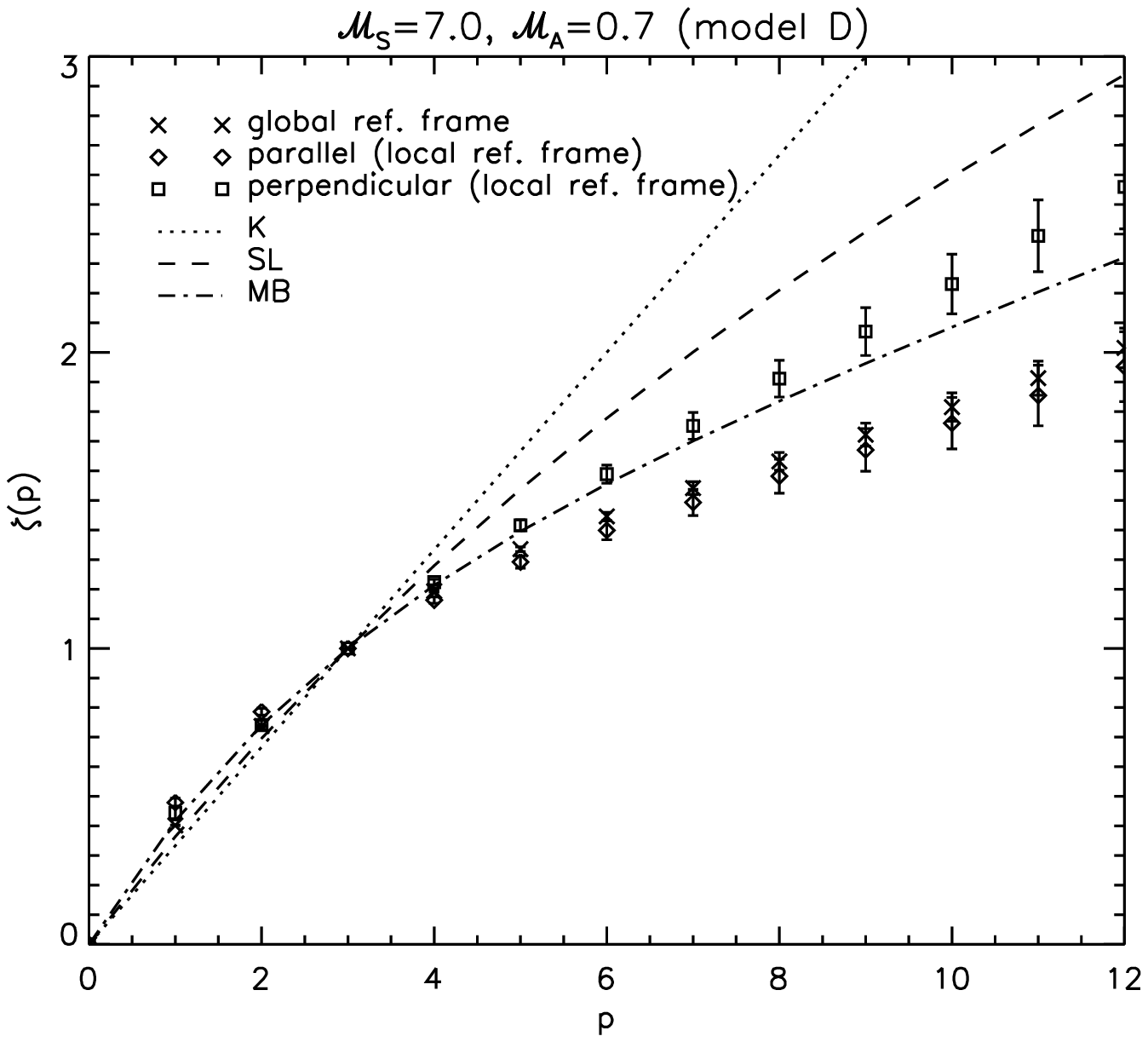}
  \caption{ Intermittency of velocities. K-no intermittency,
SL- She-Leveque model with dissipation in 1D structures,
MB- She-Leveque model with dissipation in 2D structures.
{\it left panel}: Intermittency  
        exponents for Alfvenic modes.  
 {\it central panel}: Intermittency 
        exponents for slow modes. 
{\it right panel}: Intermittency exponents for
fast modes. From Kowal \& Lazarian 2006.
}
\label{fig_coupling}
\end{figure*}

Our study in Kowal \& Lazarian (2006) showed that the intermittency of
Alfven modes, as we expected, do not change appreciably with Mach number.
The corresponding dimension of the dissipation structures is
close to one, if the calculations are done in terms of velocities and
local system of reference is used. 
The change in the intermittencies of slow and fast modes is more pronounced.
The corresponding plots are given in Fig.~3.

\section{What is the intermittency of density?}

Density fluctuations depend on the Mach number of turbulence. For
high Mach numbers the density spectrum gets shallow which is related
to the clumps and density sheets created by shocks (see Padoan et al. 2004,
Cho \& Lazarian 2004). These shocks constitute a small fraction of volume,
while in the bulk of the volume the shearing motions affect the density
structure. In Beresnyak, Lazarian \& Cho (2005) it was shown that the
logarithm of density exhibits both GS95 spectrum and anisotropy\footnote{We
note that the density itself does not show anisotropy, which means
that the large density peaks that scale with the squared of the
Mach number, are not much affected  by magnetic field.} (see Fig.~4).
 Our 
intermittency study in Kowal \& Lazarian (2006) showed that the fluctuations
of $\log\rho$ are much more regular than the fluctuations of $\rho$. The
intermittency of $\log \rho$ is somewhat higher than that of velocity (Fig.~5).
Nevertheless, the remarkable regularity of $\log \rho$ should have its
physical explanation. A possible one is related to the multiplicative
symmetry with respect to density in the equations for isothermal
hydro (see Passot \& Vazquez-Semadeni 1998). This means that if a
stochastic process disturbs the density, it results in perturbations
of density being multiplied rather than added together. Consequently,
the distribution of density for a Gaussian driving of turbulence tends to be
lognormal, which is consistent with the probability distribution
function measurements in
 Beresnyak, Lazarian \& Cho (2005). In Kowal \& Lazarian  (2006) we
used a {\it genus} measure (see Gott et al. 1990) and confirmed that
$\log  \rho$ is close to Gaussian in our MHD simulations.

\begin{figure*}
  \includegraphics[width=0.3\textwidth]{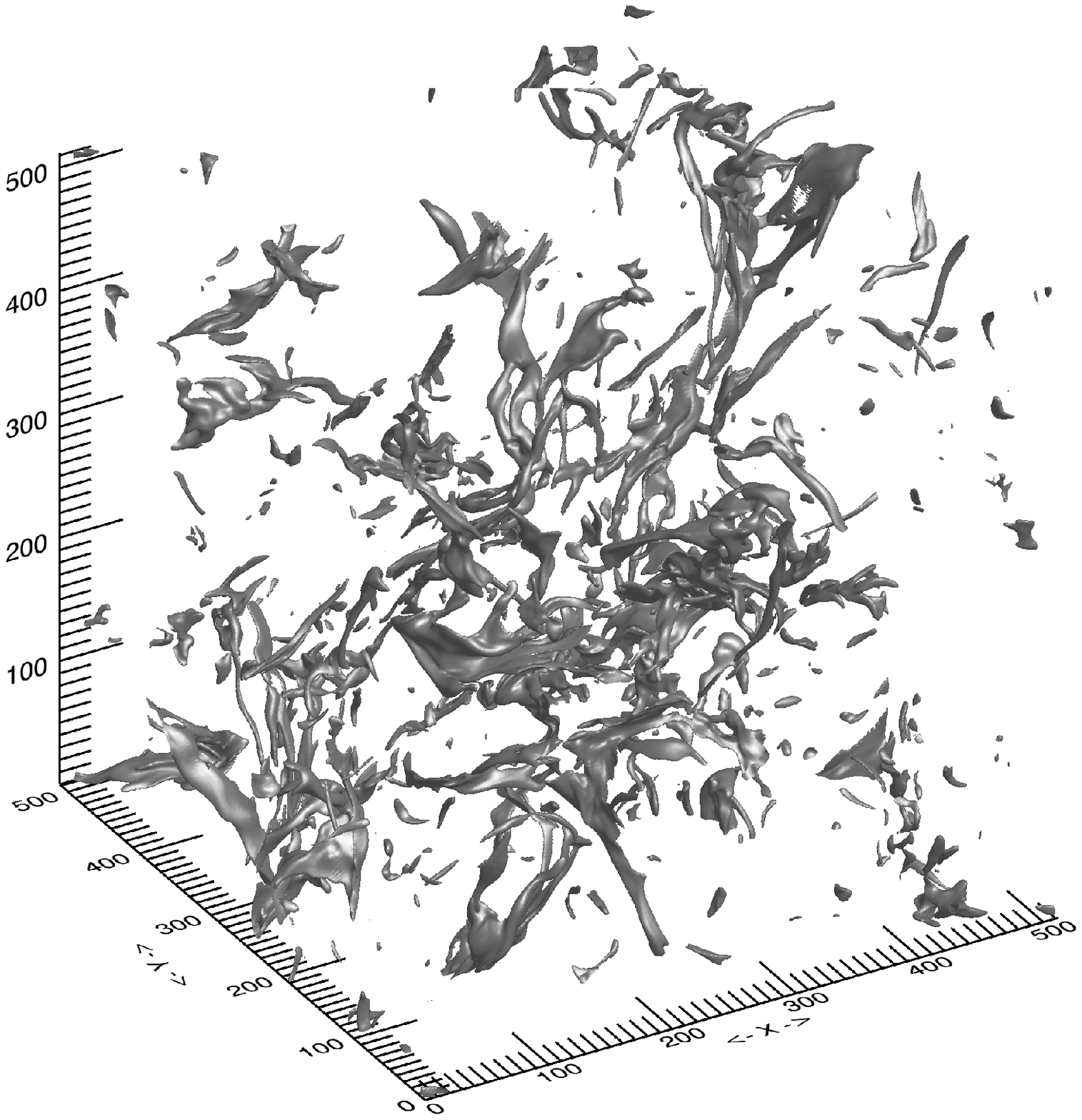}
\hfill
  \includegraphics[width=0.3\textwidth]{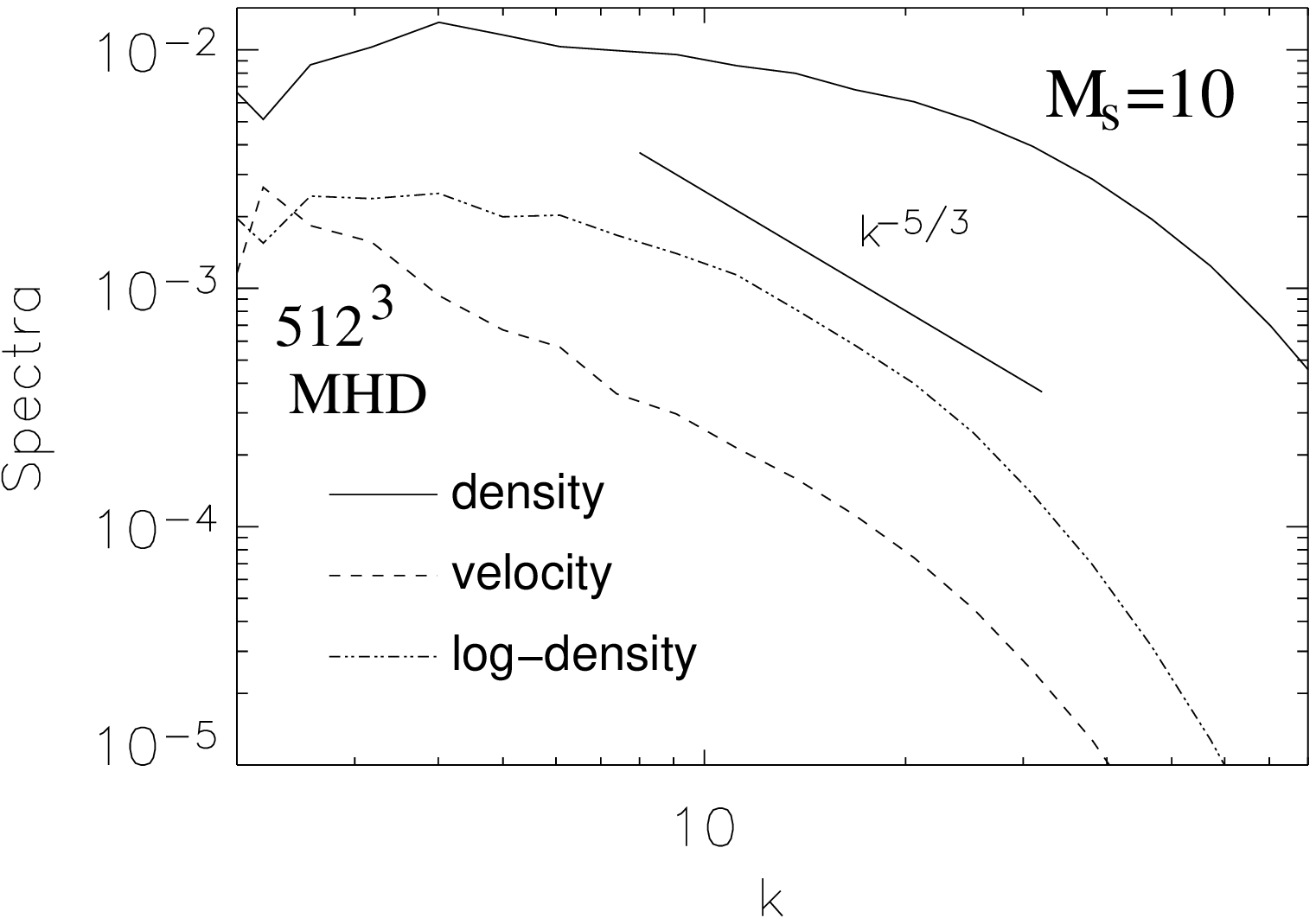}
\hfill
  \includegraphics[width=0.3\textwidth]{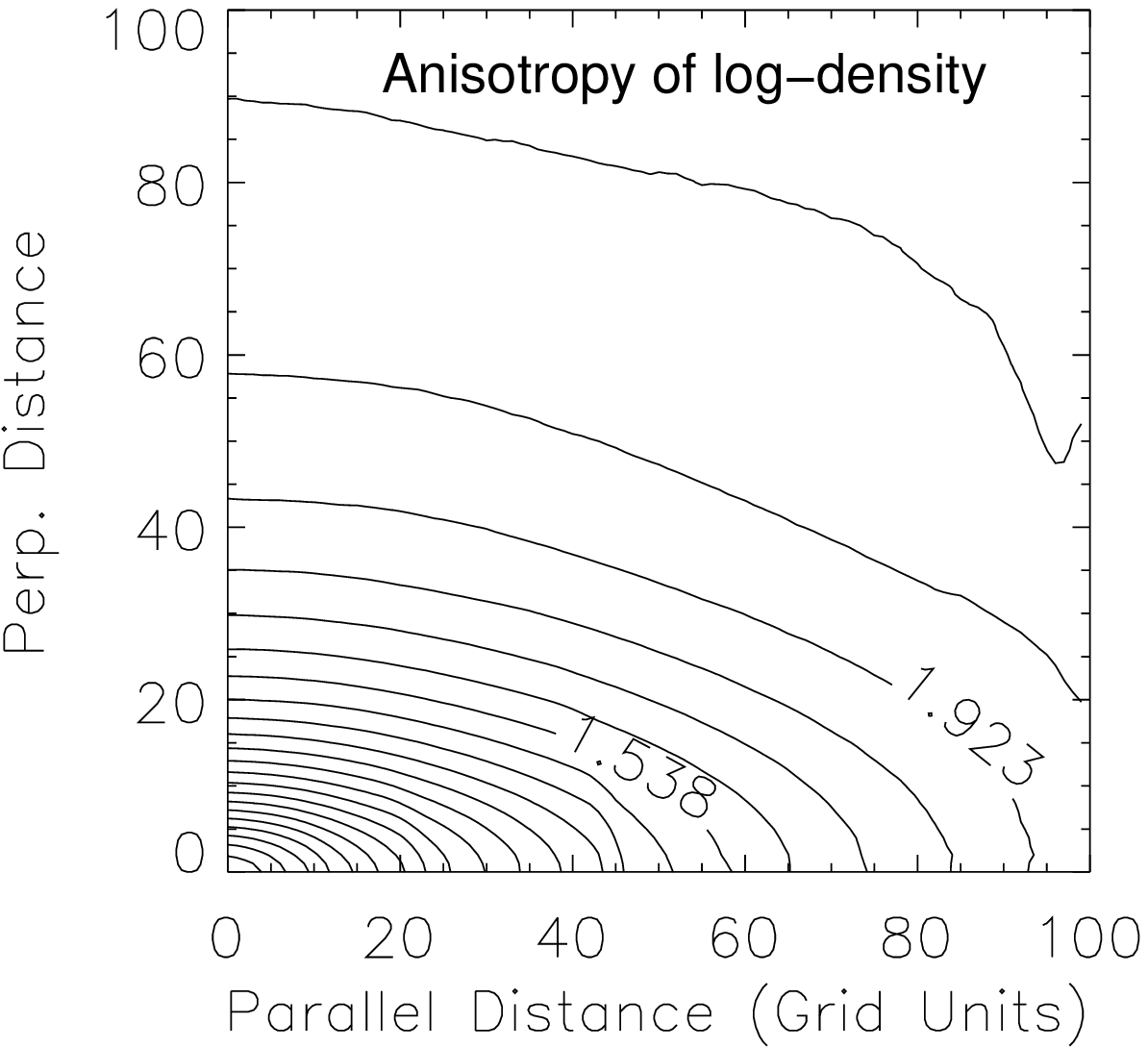}
  \caption{{\it Left}: The isosurfaces of density, corresponding to 10 mean
densities. {\it Middle}: Spectra of log  density tends to be Kolmogorov,
while the density at high Mach number is shallow. {\it Right} The contours
of isocorrelation of $\log\rho$ are very similar to  those for
velocity in GS95 picture. By Beresnyak \& Lazarian.
}
\label{fig_log}
\end{figure*}

Magnetic forces should affect the multiplicative symmetry above. 
However, they do not affect the compressions of gas parallel to magnetic
field. Those compressions in magnetically dominated fluid will be of highest
intensity and therefore most important. They are sheared by Alfvenic
modes, as their own evolution will be slower than that imposed by the
Alfvenic cascade (see theoretical considerations in GS95, CL03). We note that the shearing does not affect the pdfs, but it
does affects the spectra and anisotropy of the turbulence. This explains
close relationship between the properties of velocity in our simulations
and those of density.

\begin{figure*}
  \includegraphics[width=0.3\textwidth]{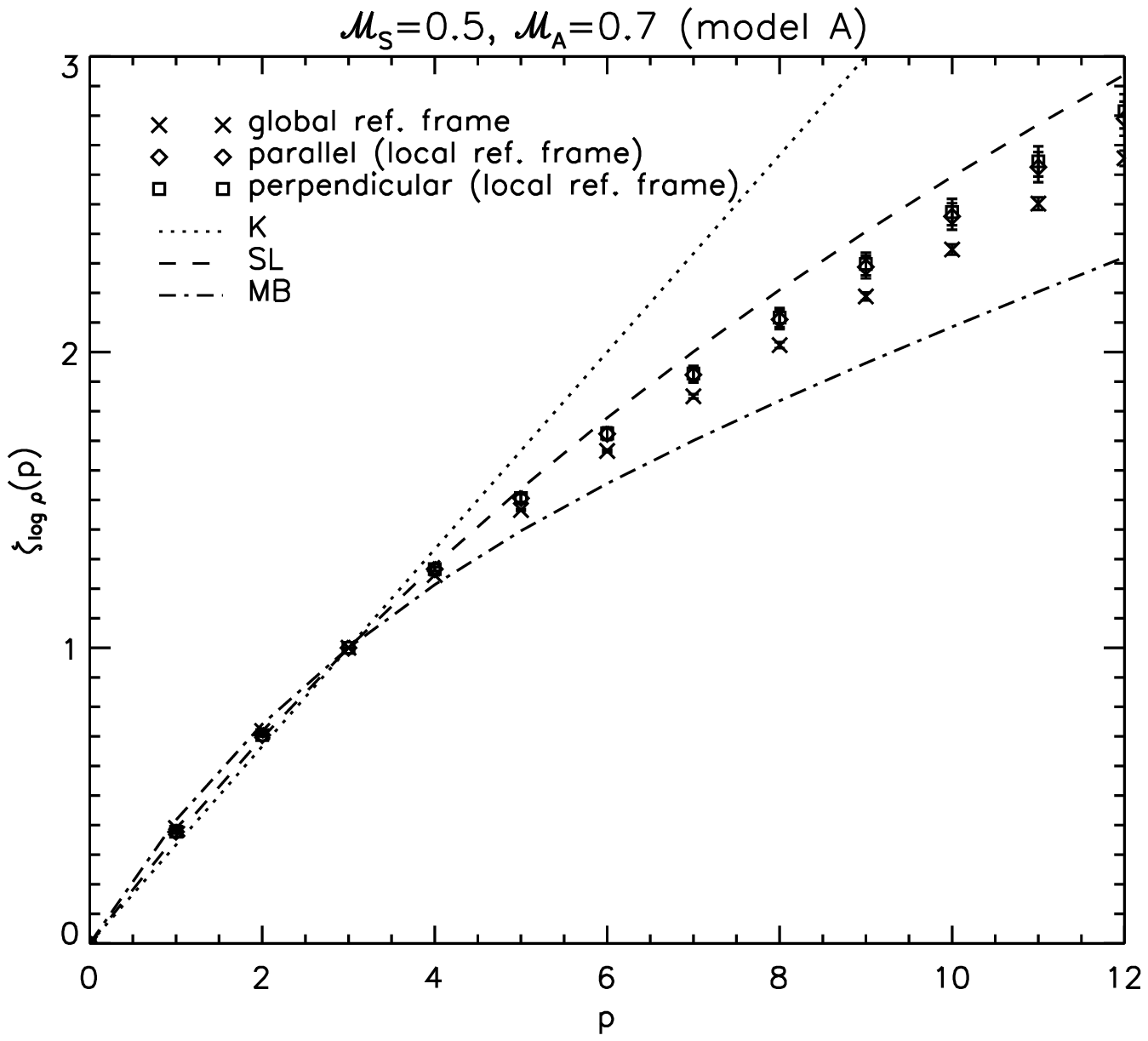}
\hfill
  \includegraphics[width=0.3\textwidth]{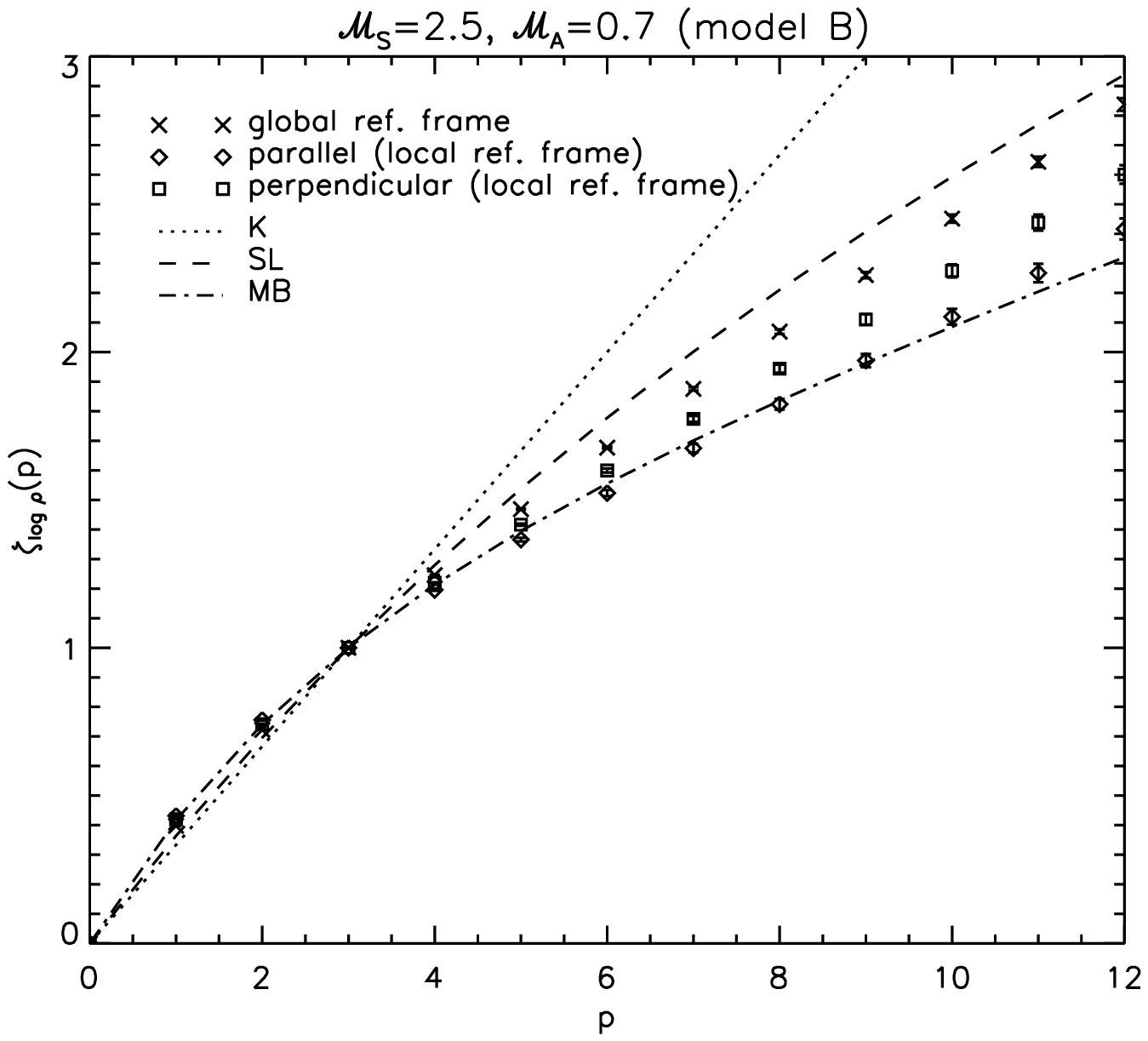}
\hfill
  \includegraphics[width=0.3\textwidth]{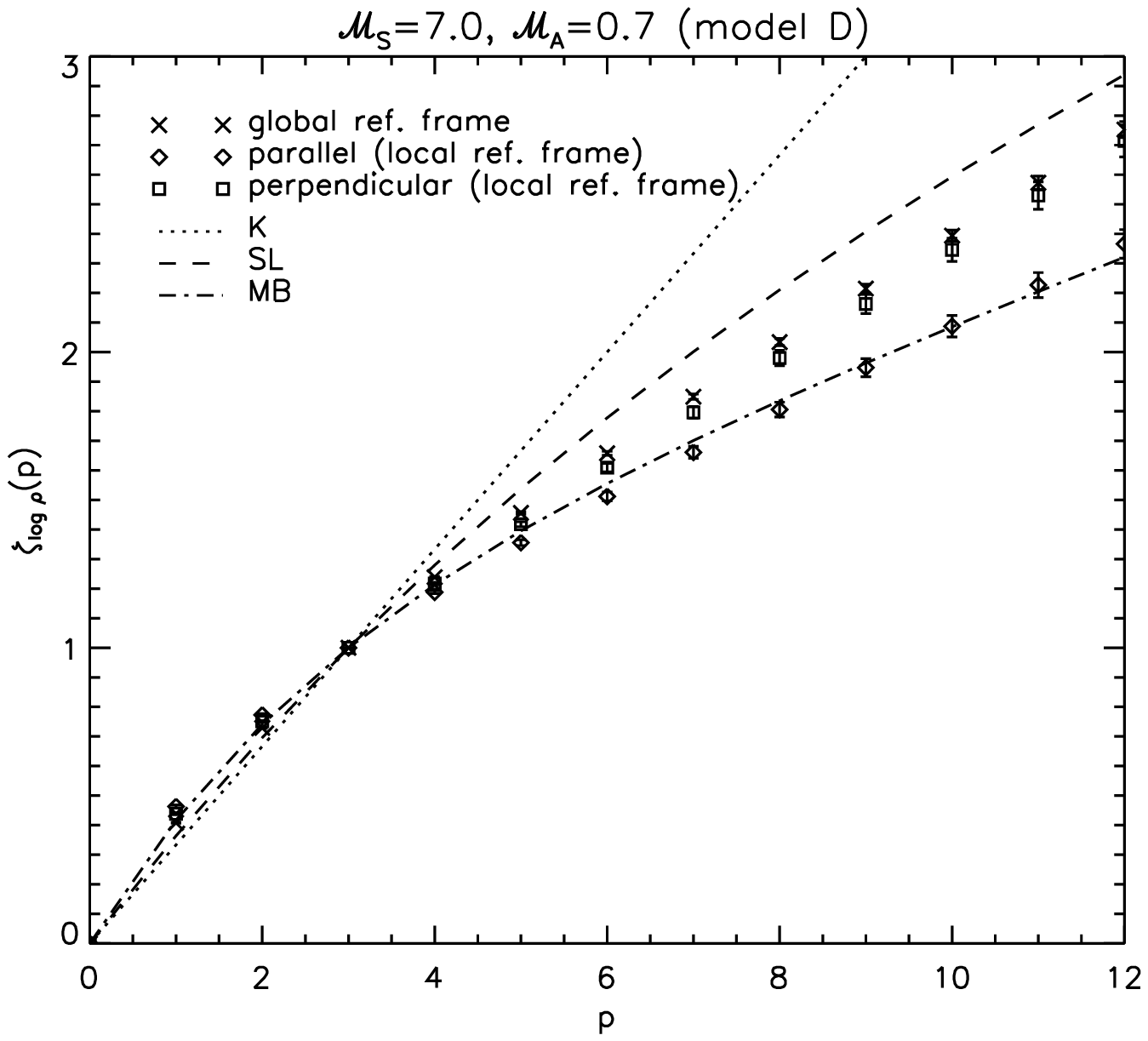}
  \caption{ Intermittency of density for
different models computed. K-no intermittency,
SL- She-Leveque model with dissipation in 1D structures,
MB- She-Leveque model with dissipation in 2D structures.
 From Kowal \& Lazarian 2006.
}
\label{fig_density}
\end{figure*}

The advantages of using  $\log\rho$ rather than $\rho$ for studies of 
supersonic turbulence were first realized in Porter et al. (1998ab).
However, important attempts to study the intermittency of density rather than
its logarithm were made in Boldyrev et al. (2002) and Padoan et al. (2003).

\section{What is the intermittency of MHD turbulence in
the viscosity-damped regime?}

Viscosity may be higher than resistivity, for instance, in a partially
ionized gas. The neutrals do not follow magnetic field lines and thus
produce viscosity, while their effect to conductivity is less pronounced.

In hydro turbulence viscosity sets a viscous scale with motions on smaller
scales being exponentially suppressed. This is natural, as at the viscous
scale the kinetic energy is dissipated rather than being transferred further. This means the end of hydro cascade, but the MHD turbulence does
evolve below the viscous scale, provided that the resistivity is lower than
viscosity. Indeed, magnetic fields can be stretched by turbulent motions
at larger scales, with the shear from the eddies at the damping scale
being most important. This means a new, viscosity-damped regime of
MHD turbulence (Cho, Lazarian, \& Vishniac 2002b).

A theoretical model for this regime of turbulence is given in Lazarian,
Vishniac, \& Cho (2004, henceforth LVC04). There it is shown that both velocities and densities
below the viscous scale form at a scale $l$
intermittent structures with  the scale-dependent
filling factor $\phi_l$, $v_l^2=\phi_l \hat{v_l}$, $b_l^2=\phi_l \hat{b_l}$,
where $\hat{v_l}$ and $\hat{b_l}$ correspond to, respectively, velocity
and magnetic field within the subvolumes. The cascading happens
on the turnover time for the eddies at the viscous scale, which
means $b_l^2\sim const$. For the incompressible fluid
LVC04 predicts that the filling factor $\phi_l$ is proportional to $l$;
the velocity spectrum {\it below} viscosity cutoff scales as $k^{-4}$,
while the magnetic field spectrum scales as $k^{-1}$. This all is 
consistent with the incompressible 
numerical simulations in CLV03, but the theory may need to be modified for the
compressible medium, where the magnetic filaments are allowed to expand.

\begin{figure*}
  \includegraphics[width=0.3\textwidth]{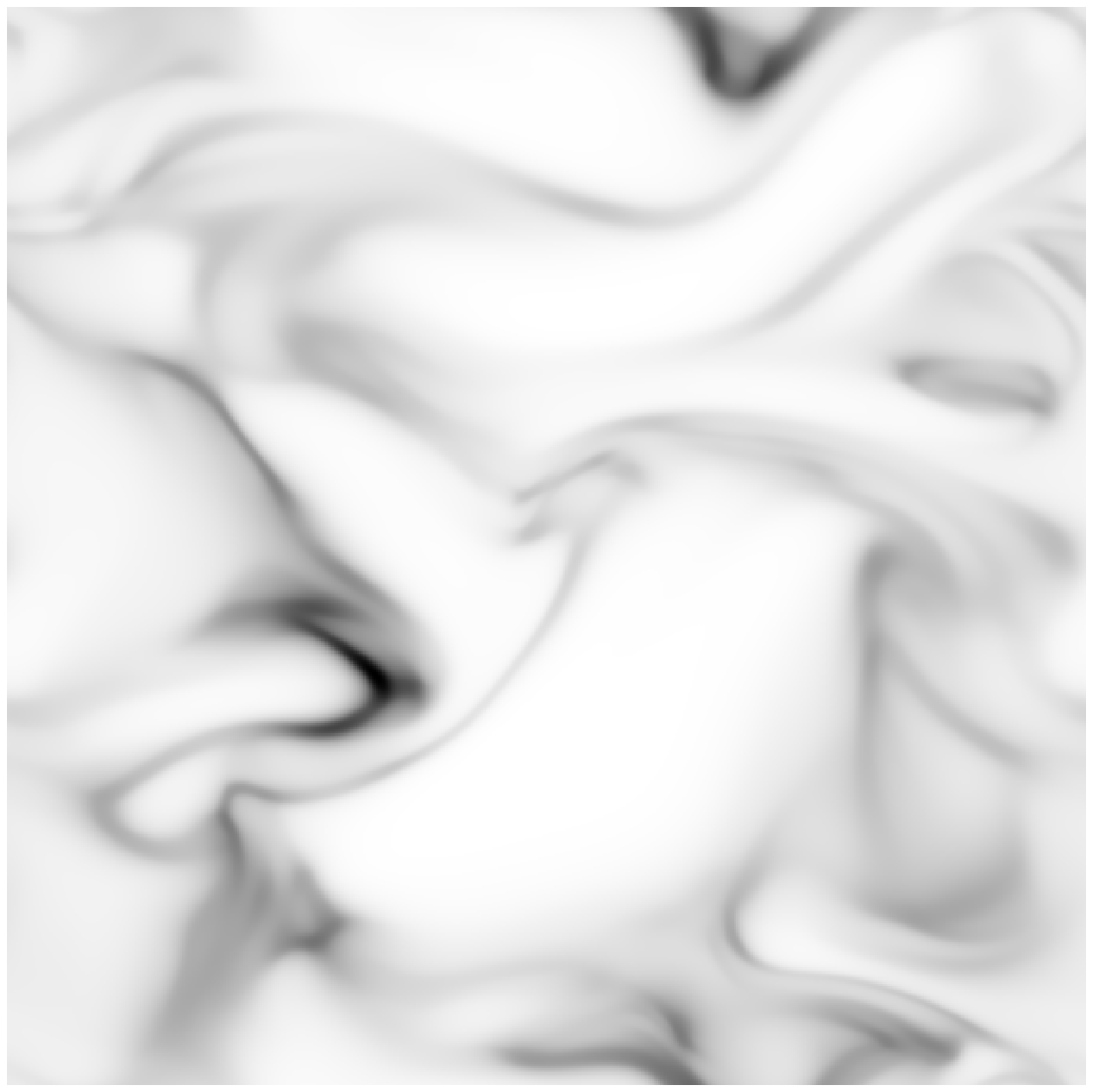}
\hfill
  \includegraphics[width=0.3\textwidth]{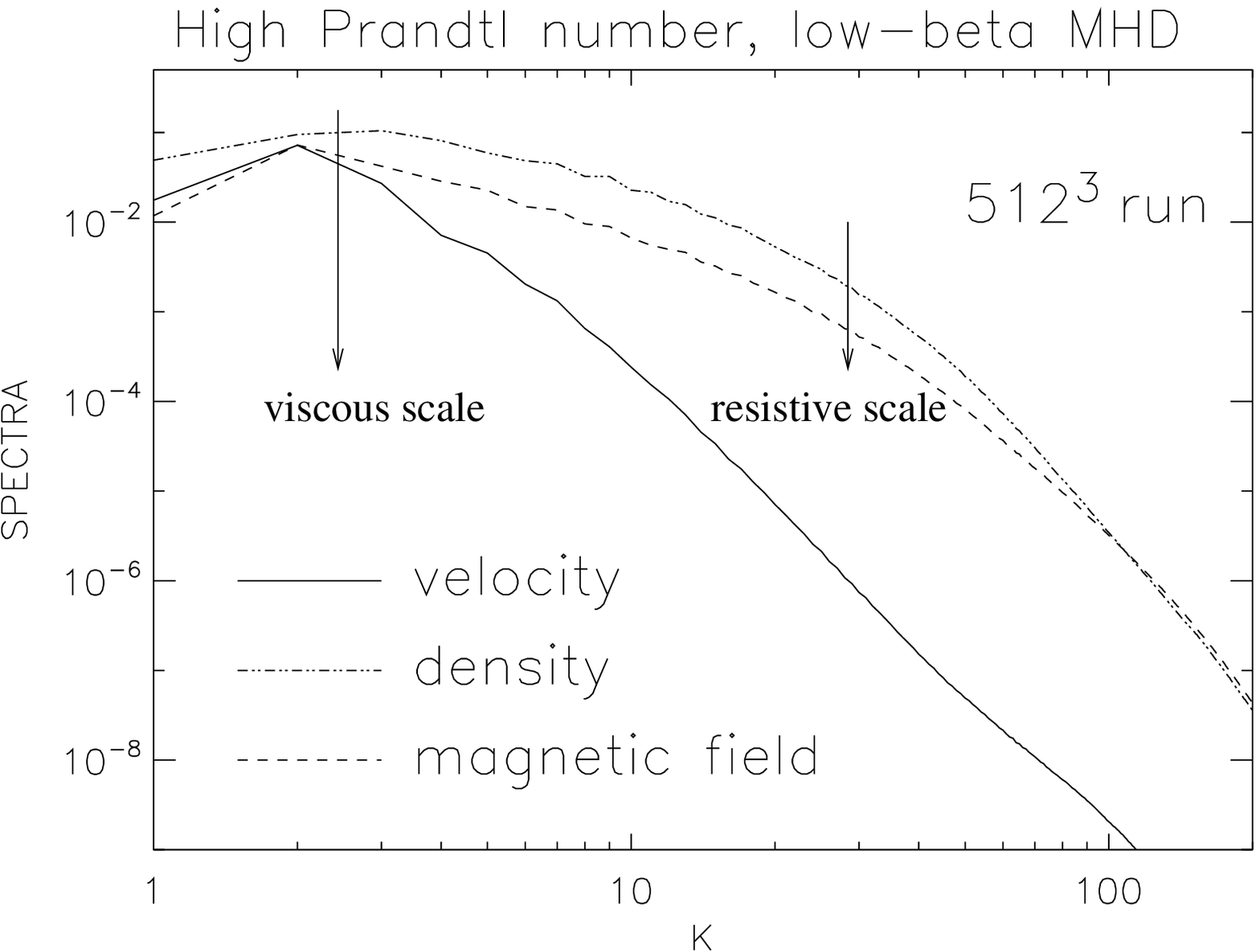}
\hfill
  \includegraphics[width=0.3\textwidth]{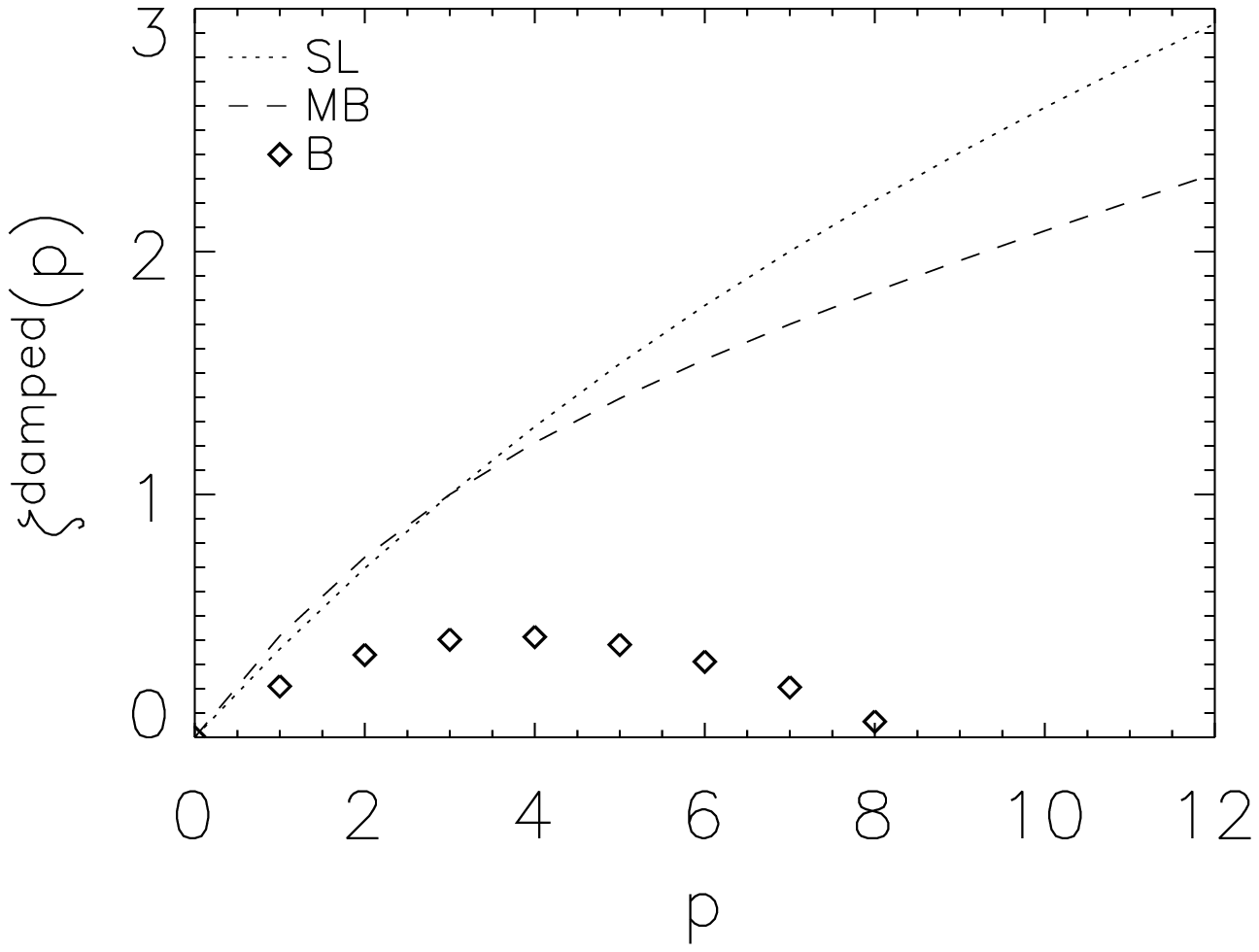}
  \caption{{\it Left}: 
Filaments of density created by magnetic compression in the
slice of data cube of the viscosity-damped regime of MHD turbulence. 
{\it Middle}: Spectra of density and magnetic field are similar,
while the velocity is damped (as a steep power law). Note that
in this regime the resistive scale is $L Rm^{-1/2}$. {\it Right}:
Intermittency spectral index for the incompressible MHD turbulence
in the viscosity-damped regime.
}
\label{fig_damped}
\end{figure*}

For the extreme intermittency of the magnetic field suggested in LVC04
the higher moments of structure functions 
$S_p\sim \hat{b_l}^p \phi_l$ which means that $S_p\sim l^{1-p/2}$. 
The concentration of magnetic field in thin filaments gives rise to 
resistive loses that should eventually make  $\xi_p=0$ for sufficiently
large $p$. In Fig.~6 we see this general tendency for high $p$.
For the absence of the more precise correspondence we may blame
(a) our crude model for estimating $\xi$, (b) numerical effects, and (c)
LVC04 model itself. Addressing the issue (b), we would say that the 
compelling arguments in the model provide $k^{-1}$ spectrum and this
would provide $\xi(2)=0$ in accordance with the intermittency model
above. However, due to numerical effects identified in LVC04
the spectrum of magnetic fluctuations is slightly steeper.

\section{What is the intermittency in polarization?}

Polarization is a more subtle effect that is observed in 
Alfvenic turbulence.
Since Alfven waves are transversal, i.e. the velocities are perpendicular
to both ${\bf B}$ and ${\bf k}$, they have two polarizations. It can be
shown that only waves with orthogonal polarizations efficiently interact 
with each other (see Biskamp 2003). Consider two  oppositely moving wave
packets of the {\it same} energy, partially polarized in the same 
direction.
Their interaction can be roughly understood as the interaction of 2 pairs of
orthogonally polarized wave packets (i.e. 4 wave packets total). 
Within each of these packets there is
{\it imbalance}, i.e. the energy in one of the interacting waves 
is larger than
the energy of the other wave moving in the opposite direction. Studies
of the imbalanced turbulence 
(see Biskamp 2003, Cho, Lazarian \& Vishniac 2002, Boldyrev 2005 and ref. therein) 
show that the 
stronger component induces a faster decay of the weaker one (it shears it
more). As the result, the polarizations of the wave packets increase and the
rate of turbulence decay decreases. Note, that the increase of degree of
polarization was observed in the simulations by
 Maron \& Goldreich (2001, henceforth MG01).

The recent upsurge of interest to effects of polarization is related to
the ongoing attempts to understand the actual slope of the MHD turbulence
spectrum. As the numerical simulations obtained a sufficient inertial
range, it became possible to distinguish the spectral slopes $-5/3$ and $-3/2$. In particular, Boldyrev (2005, 2006) associated the $-3/2$ slope 
measured in MHD simulations of low Alfven Mach number ($M_A\ll  1$) turbulence
(MG01, Muller \& Grappin 2005) with the additional
asymmetries of turbulent eddies in the plane perpendicular to magnetic
field. This was in contrast to a remark in MG01
that related the flattening of the spectrum to the turbulence 
intermittency. 

\begin{figure*}
  \includegraphics[width=0.32\textwidth]{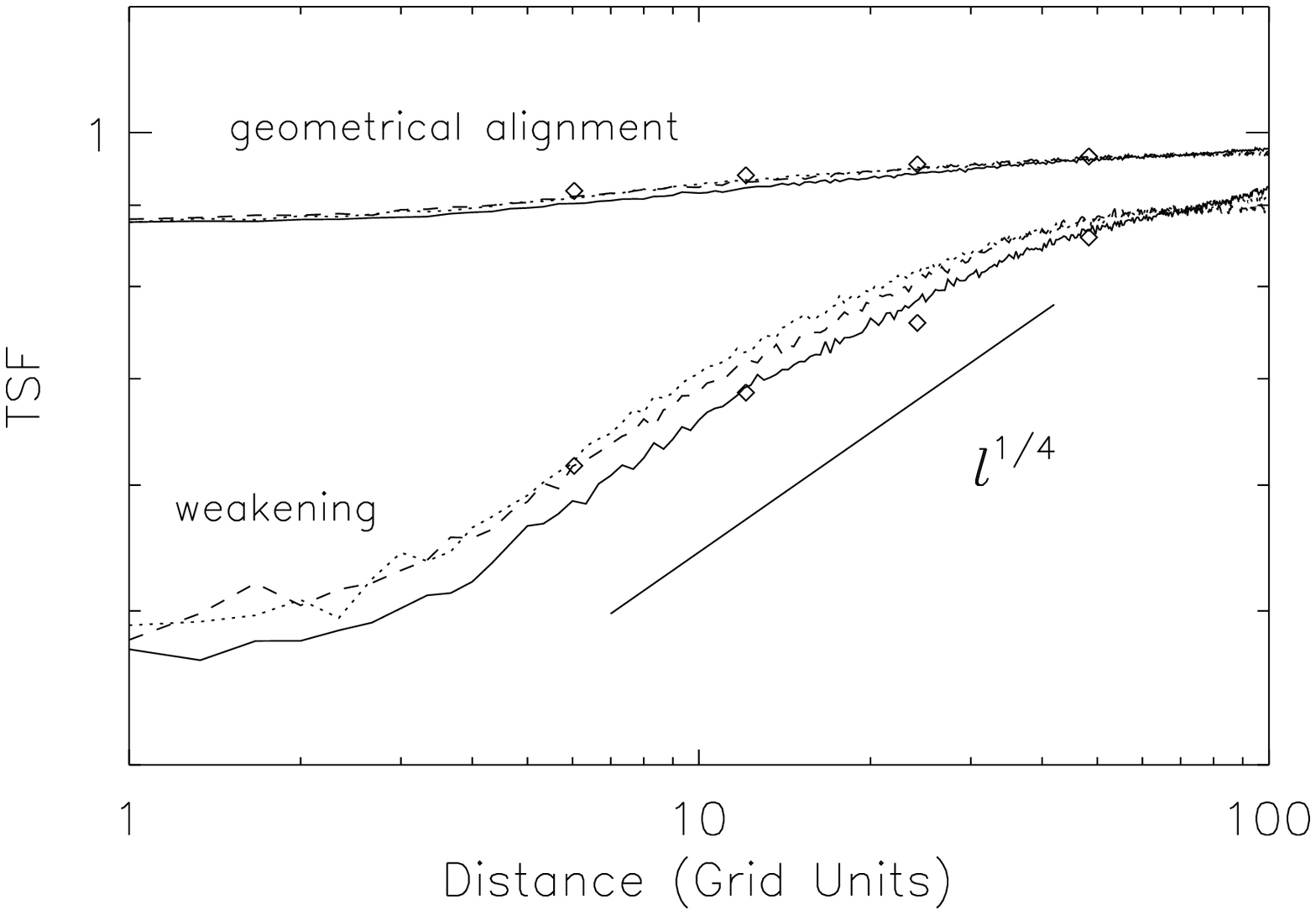}
  \hfill
  \includegraphics[width=0.32\textwidth]{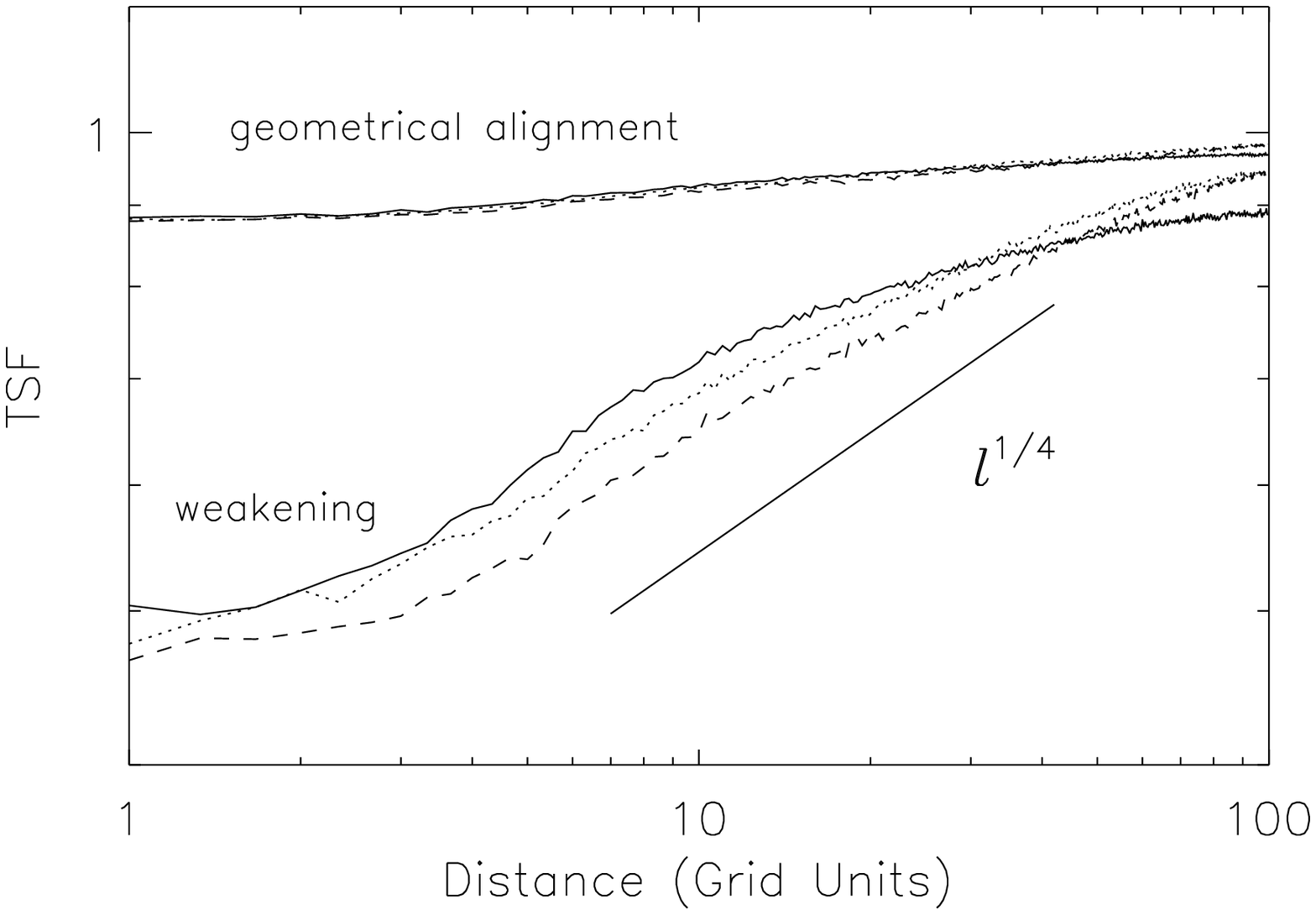}
  \hfill
  \includegraphics[width=0.32\textwidth]{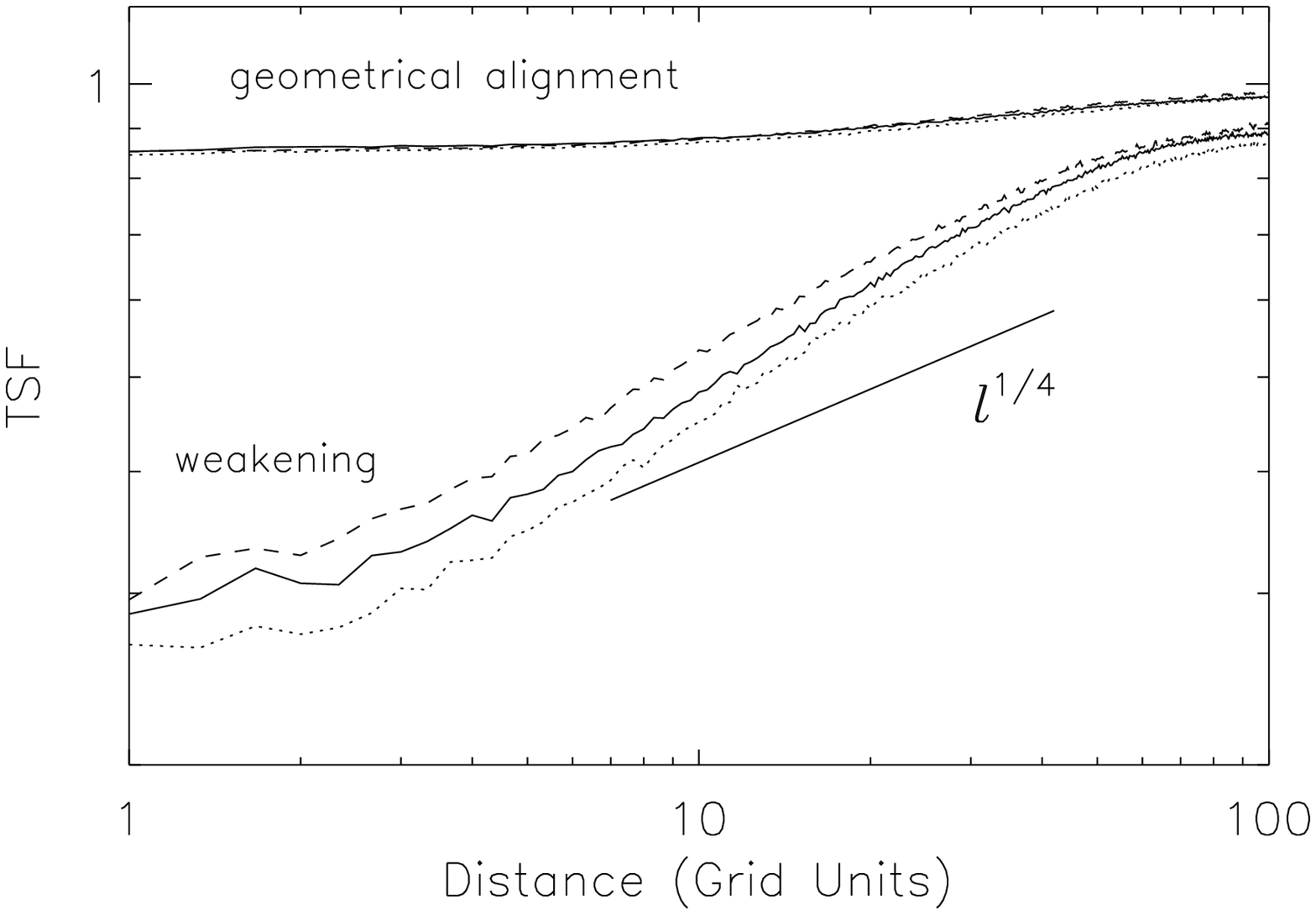}
  \caption{ The geometrical alignment (upper curve) and the weakening of
 interaction factor (lower curve), {\it left panel}: incompressible, $M_A=0.7$,
 Solid, dotted and dashed lines are for three simulations,
separated by Alfv\'enic cross time. {\it Central panel}: incompressible, $M_A=1.0$,
same notation. {\it Right panel}: Compressible, $M_A=1$, $M_s=1$,
we used a global mean magnetic field for both mode separation and scale separation. Clearly in all these cases pure geometric alignment provides
rather marginal effect. The weakening of the interaction arises from
intermittent regions with high polarization and high wave amplitudes.}
\label{SF}
\end{figure*}

Our study in Beresnyak \& Lazarian (2006) shows that the 
{\it polarization
 intermittency} plays an important role in MHD turbulence. 
We found that the MHD turbulence volume spontaneously develops
regions that are characterized by both high amplitude of
fluctuations and high degree of polarization (see Figure 7).
At the same time 
the volume-averaged pure geometrical alignment of the oppositely
moving wave  packets is rather marginal for our simulations.  
The degree of the 
spontaneous polarization and the corresponding weakening of the cascade
increases with the decrease of the scale. As the result one should
expect the turbulence spectrum to get flatter.

An additional interesting effect noticed in Beresnyak \& Lazarian (2006)
is that the intermittency of the nonlinear term constructed with
the fluctuations of the Elsasser variables $u$ and $w$ multiplied by 
$\sin \theta$, where $\theta$ is the angle between 
${\bf w}$ and ${\bf  u}$ is less intermittent than product $u$ and $w$
which can be interpreted as the tendency of turbulence to 
``self-organize'' to provide a more homogeneous dissipation. Interestingly enough, the nonlinear term demonstrated much better extendended
self-similarity than the $u$ and $w$ product. This indicates a rather
fundamental nature of the nonlinear term. Naturally, more studies are
required for this opening field.

\section{What is the future of the field?}

Extended self-similarity allows to perform informative intermittency studies
using computers. A substantial impetus can be obtained from observational
studies. Such studies have been performed already (see Falgarone et al. 2005,
Padoan et al. 2003), but velocity statistics is still illusive. A substantial
progress in understanding of when the observations reflect the statistics 
of velocity (see
 Esquivel \& Lazarian 2005 and references therein) allows us to hope that measures
of the intermittency of interstellar velocities will be available soon.
This all should allow predictions of the intermittency theory to be tested
with observations. A developed intermittency theory should initiate a chain
reaction of changes within different branches of astrophysics, including
the transport theory, acceleration of cosmic rays and, possibly, even
astrochemistry. It also should contribute to a better interpretation
of the simulation results in terms of astrophysical phenomena and may
eventually results in new computational tools that would adequately represent
the turbulence-related astrophysical processes at any scale of interest.

{\bf Acknowledgments}{I thank Ethan Vishniac, Jungyeon Cho and Andrey
Beresnyak 
for valuable discussions of turbulence intermittency and effects of
polarization on the Alfvenic cascade. The work is supported by
the NSF Grant AST-0307869 and the NSF Center for Magnetic 
Self-Organization in Laboratory and Astrophysical Plasmas}.

\end{article}
\end{document}